\documentclass[journal]{IEEEtran}

\usepackage{graphicx}
\usepackage{url}
\usepackage{hyperref}

\expandafter\let\csname equation*\endcsname\relax

\expandafter\let\csname endequation*\endcsname\relax
\usepackage{amsmath}

\usepackage{subfigure}
\usepackage[capitalise]{cleveref}

\DeclareMathOperator*{\argmin}{arg\,min}
\newcommand{\eg}{e.g., }
\newcommand{\ie}{i.e., }

\begin{document}
\title{Analytical Estimation of Beamforming\\ Speed-of-Sound Using Transmission Geometry}
\author{\IEEEauthorblockN{
Can Deniz Bezek,
Orcun Goksel\\
\IEEEauthorblockA{
Department of Information Technology, Uppsala University, Sweden
}
\thanks{Funding was provided by the Medtech Science and Innovation Centre at Uppsala University.}
\thanks{$\{$can.deniz.bezek,orcun.goksel$\}$@it.uu.se}}
}

\maketitle

\begin{abstract}
Most ultrasound imaging techniques necessitate the fundamental step of converting temporal signals received from transducer elements into a spatial echogenecity map.
This beamforming (BF) step requires the knowledge of speed-of-sound (SoS) value in the imaged medium. 
An incorrect assumption of BF SoS leads to aberration artifacts, not only deteriorating the quality and resolution of conventional brightness mode (B-mode) images, hence limiting their clinical usability, but also impairing other ultrasound modalities such as elastography and spatial SoS reconstructions, which rely on faithfully beamformed images as their input. 
In this work, we propose an analytical method for estimating BF SoS.
We show that pixel-wise relative shifts between frames beamformed with an assumed SoS is a function of geometric disparities of the transmission paths and the error in such SoS assumption.
Using this relation, we devise an analytical model, the closed form solution of which yields the difference between the assumed and the true SoS in the medium. 
Based on this, we correct the BF SoS, which can also be applied iteratively.
Both in simulations and experiments, lateral B-mode resolution is shown to be improved by $\approx$25\% compared to that with an initial SoS assumption error of 3.3\%~(50\,m/s), while localization artifacts from beamforming are also corrected.
After 5 iterations, our method achieves BF SoS errors of under 0.6\,m/s in simulations and under 1\,m/s in experimental phantom data.
Residual time-delay errors in beamforming 32 numerical phantoms are shown to reduce down to 0.07\,$\mathbf{\mu}$s, with average improvements of up to 21 folds compared to initial inaccurate assumptions.
We additionally show the utility of the proposed method in imaging local SoS maps, where using our correction method reduces reconstruction root-mean-square errors substantially, down to their lower-bound with actual BF SoS.  
\end{abstract}

\section{Introduction}
Speed-of-Sound (SoS) is the longitudinal travel rate of acoustic waves within a medium. 
Most ultrasound (US) imaging techniques in their image formation process require the prior knowledge of the medium SoS.
Since this value in general is not known for a target medium being imaged, assumed generic, spatially-constant SoS values are used as an approximation.
For example, B-mode imaging during a liver US exam typically utilizes an average liver SoS value from the literature.
However in reality, SoS may change largely depending on the imaged medium; \eg SoS values for muscles ($1585$\,m/s) and fat ($1440$\,m/s) may differ up to 10\% \cite{szabo2004diagnostic}.
Even the same anatomical structure may show large variations across the population; \eg varying breast SoS values were reported with an ultrasound computed tomography (USCT) setup in \cite{duric2007detection, duric2013breast, o2017ultrasound},
breast SoS variations of up to 5.6\% were reported in a study with over 100 patients using a hand-held ultrasound system in \cite{Sanabria_breast-density_18}, and a 2\% SoS difference was reported between calf muscles of young and elderly females in \cite{sanabria_sacropenia_2019}.
Thus, in many cases an assumed SoS value is not ideal for optimal imaging results.

Beamforming is the projection of temporal signals received by an US transducer into a spatial US image, which requires an assumed SoS value of the tissue.
Any discrepancy between the actual and assumed SoS values in beamforming leads to aberration artifacts; causing blurring and reduced resolution of common brightness mode (B-mode) images, hence limiting their diagnostic usability. 
Besides the perceptual deterioration of B-mode images, other ultrasound modalities including quantitative ultrasound techniques that depend on beamformed images may also be hindered by suboptimal beamforming.
For instance, in \cite{Chintada_phase-aberration_21} shear-wave elastography (SWE) measurements were shown to vary largely based on the SoS assumption during beamforming.
Tomographic reconstruction of local SoS maps \cite{rau_2021} is also hindered by inaccuracies in global SoS assumption in beamforming, as was demonstrated in \cite{xenia_2021}.
Therefore, utilizing a correct SoS assumption is of great importance in multiple US image formation pipelines, often through the utilized beamforming mechanism, and hence SoS estimation has been studied by several groups.
These works estimate either a single (homogeneous) global SoS value or a spatial (heterogeneous) SoS map. 

Among single value estimation methods, a least-squares fitting of a $2^{nd}$-order polynomial to the echo profile was proposed in \cite{anderson_1998} for estimating a global SoS value. 
This work was extended in \cite{byram_2012} for measuring SoS in layered phantoms. 
In \cite{ophir1986}, a beam tracking method is used to estimate SoS along an arbitrary line in tissue.
In \cite{krucker2004}, SoS is found based on registering multiple electronically steered ultrasound images. 
Several approaches ave been proposed to find an optimal SoS value by maximizing a specific quality metric, for instance based on spatial frequency content analysis at pre-selected image region \cite{Napolitano2006}, echo deconvolution via estimated point spread functions \cite{shin2010}, via speckle analysis \cite{qu2012}, via minimal average instantaneous phase variance of channel RF data \cite{yoon2011} and its extension by phase dispersion also considering beamformed intensities \cite{perrot2021}. 
An average tissue SoS value can be measured from the known or calibrated distance to an acoustic reflector, as demonstrated in \cite{Sanabria_breast-density_18} for breast density prediction and in \cite{Ruby_quantification_21} for muscular degeneration. 
In \cite{shen_2020} a mean SoS value was estimated using the signal coherence between different transmit (Tx) and receive (Rx) paths in multi-angle plane wave imaging.
In \cite{xenia_2021}, mean SoS was estimated based geometric disparities between different transducer elements on lateral disparity profiles at chosen depths, which were mapped to correction values based on calibrations from prior simulations.
Such calibrations are tedious and error-prone, and this geometric model makes inaccurate approximations.

For heterogeneous mapping of SoS, steered plane waves with a frequency domain reconstruction was demonstrated in \cite{jaeger_2015}.
With a spatial domain reconstruction approach \cite{sanabria_2018}, SoS-based differential diagnosis of breast cancer using a hand-held US transducer was reported first time in \cite{Ruby_breast_19}.
Diverging waves were proposed in \cite{rau_2021} to minimize wavefront diffractions, thus significantly improving local SoS reconstruction results. 
The use of such tomographic SoS maps in adapting beamforming locally was shown to improve beamformed B-mode images \cite{rau_2018} and subsequently also SWE measurements \cite{Chintada_phase-aberration_21}. 
Note that all such local SoS reconstruction methods rely on displacement tracking from beamformed images, and are therefore inherently sensitive to errors from incorrect initial SoS assumption.

In this work, we develop an analytical method for the estimation of global SoS for beamforming, and present the utility of this for aberration correction in B-mode imaging as well as accuracy and robustness improvement in tomographic reconstruction of local SoS.
Preliminary results of of this work were presented in \cite{deniz2022global}.

\section{Method}
\begin{figure}
\centering
\includegraphics[width=\linewidth]{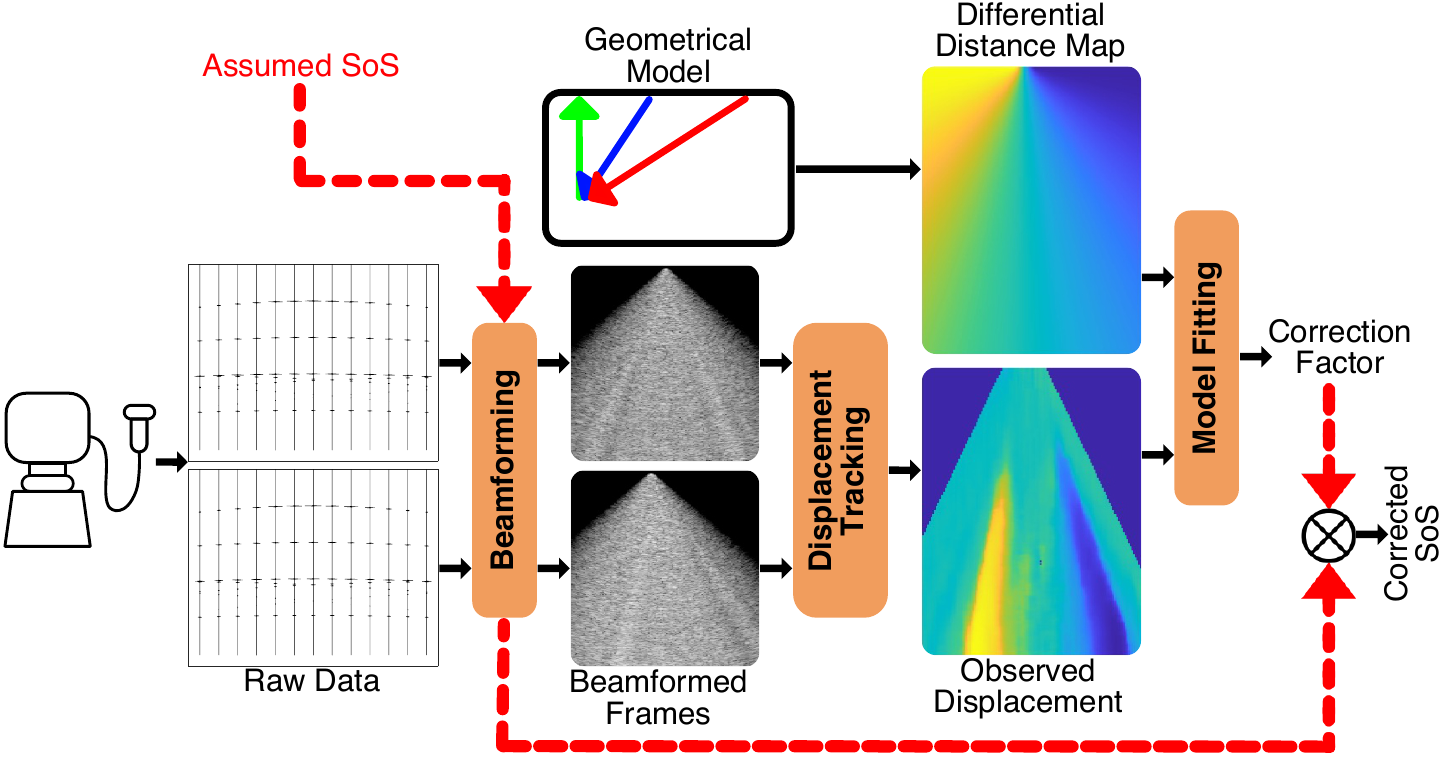}
\caption{Overview of our proposed method for estimating beamforming speed-of-sound (SoS), by correcting an initially assumed SoS value.}
\label{fig:processing_pipeline}
\end{figure}

We model the systematic offsets between geometric disparities with different Tx/Rx events, and fit this model to displacements observed between respective acquired images.
A method overview is illustrated in \cref{fig:processing_pipeline}.
First, Rx raw (temporal) data from two different Tx sequences are beamformed using an (arbitrary) assumed SoS value.
Due to the spatial distance differences between these sequences to any spatial point, any mismatch between the assumed and actual SoS values would cause a spatially-varying shift between the beamformed frames. 
Using known transducer geometry, we calculate a spatial map of expected shifts as a model for estimating any SoS mismatch.
Between the beamformed RF frames, we use standard displacement tracking algorithms to estimate the apparent shifts, which are then fitted by the precomputed model to calculate a correction factor for updating the assumed SoS value to an accurate SoS estimate. 
In contrast to a previous method \cite{xenia_2021} that rely on geometric disparities, herein we propose an accurate quantitative model and use an analytical closed-form solution, without a need for a correlative approach and calibration.

\subsection{Analytical Model}
\begin{figure}
\centering
\begin{tabular}{@{}c@{~~}c@{~~}c@{}}
\includegraphics[width=0.31\linewidth]{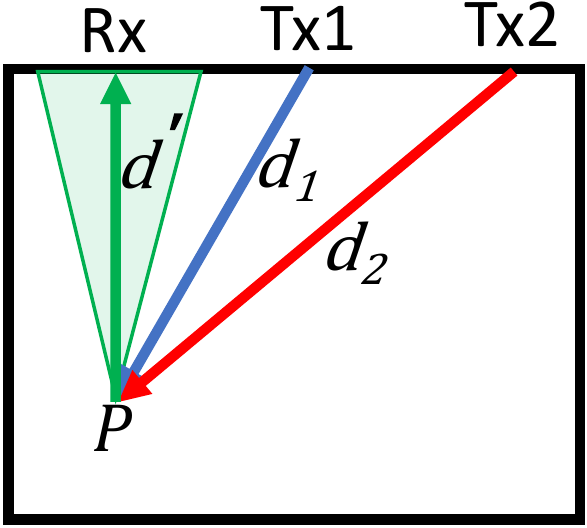}   &  \includegraphics[width=0.31\linewidth]{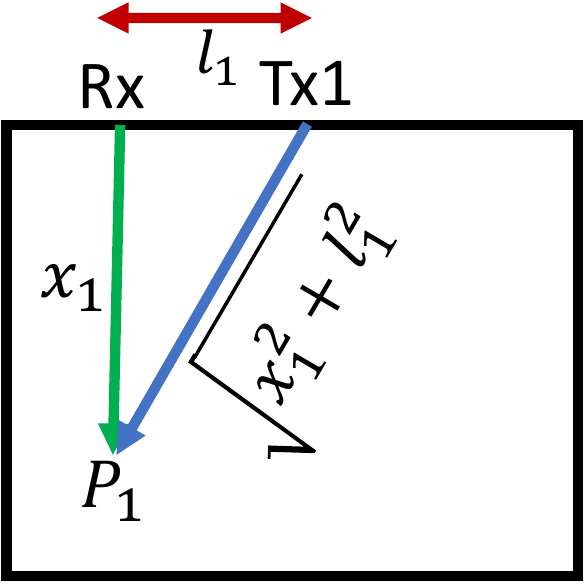}  &
\includegraphics[width=0.31\linewidth]{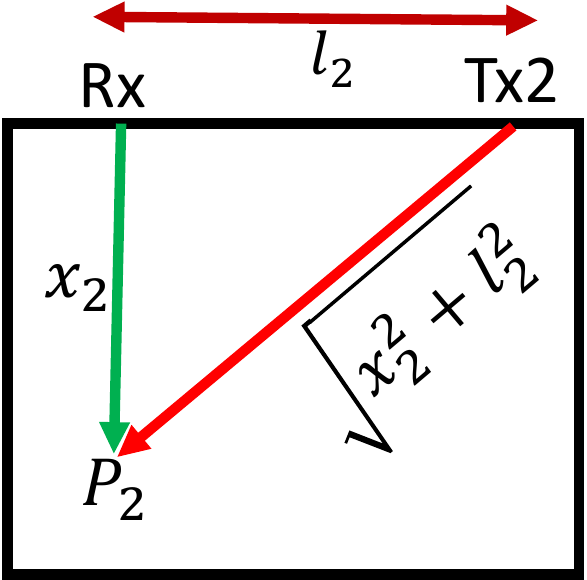} \\ 
        \footnotesize(a)~Physical scene& \footnotesize(b)~BF image from Tx1& \footnotesize(c)~BF image from Tx2
    \end{tabular}
        \caption{Illustration of our analytical model derivation, for physical tissue geometry~(a) beamformed with diverging waves from Tx1~(b) and Tx2~(c).}
        \label{fig:BF_grid_illustration}
    \end{figure}

Herein we employ a sequence with a pair of diverging wave (DW) transmits, and dynamic receive (Rx) with apertures centered over each beamformed image location.
Consider the DW transmits Tx1 and Tx2 having distances $d_1$ and $d_2$ to a fixed tissue point (feature) $P$, which is $d'$ farther from the Rx aperture, as shown in \cref{fig:BF_grid_illustration}(a).
Due to different distances and potential SoS variations in the propagation paths, the waves reflected from point $P$ arrive back at the transducer with different time-delays. 
Let these time-delays be $t_1$ for Tx1 and $t_2$ for Tx2.
These can be expressed as a sum of Tx and Rx time-delays to and from point $P$ as:
\begin{align}
t_1 &= t_\mathrm{Tx1} + t_\mathrm{Rx} = \frac{d_{1} + d^{'}}{{c_\mathrm{{}}}} \label{eq:t1}\\
t_2 &= t_\mathrm{Tx2} + t_\mathrm{Rx} = \frac{d_{2} + d^{'}}{{c_\mathrm{{}}}} \label{eq:t2}\ ,
\end{align}
where $c_\mathrm{}$ is the actual SoS in the medium.
From beamforming of the received signals from Tx1 and Tx2, let $P_1$ and $P_2$ be the respective image locations where the tissue feature $P$ appears, as seen in \cref{fig:BF_grid_illustration}(b,c). 
For convenience and to avoid repetition, the following formulations common for both transmits are written using $i$$\in$$\{1,2\}$.
Vertical distances $x_i$, between $P_i$ and the Rx aperture, are then a function of respective arrival time-delays $t_i$ as:
\begin{align}
t_i{c^\mathrm{{*}}} &= x_i + \sqrt{x_i^2 + l_i^2}\ , \label{eq:t_x_relation}
\end{align}
where $c^\mathrm{*}$ is the assumed SoS value for beamforming, and $l_i$ represents the horizontal distance from Rx to respective Tx (see~\cref{fig:BF_grid_illustration}(b,c)).
From~\eqref{eq:t_x_relation}, moving $x_i$ to the left side and taking squares yield $x_i$ in terms of the other variables as:
\begin{align}
t_i{c^\mathrm{*}}-x_i &  = \sqrt{x_i^2 + l_i^2} \\
t_i^2(c^{\mathrm{*}})^2 + x_i^2 - 2 x_it_i{c^\mathrm{*}} & = x_i^2 + l_i^2 \\
x_i & = \frac{t_i^2{(c^{\mathrm{*}})^2} - l_i^2}{2t_ic^\mathrm{*}}
\label{eq:x_expression}
\end{align}

Let us now calculate the vertical difference (spatial shift) $\Delta x$ between the appearances of points $P_i$ in their respective beamformed images, by first substituting~\eqref{eq:x_expression}, then rearranging terms and substituting~\eqref{eq:t1} and \eqref{eq:t2}, as follows: 
\begin{align}
\Delta x  & =  x_2 -  x_1\\
& = \frac{t_2^2{(c^{\mathrm{*}})^2}-l_2^2}{2t_2c^{\mathrm{*}}} - \frac{t_1^2{(c^{\mathrm{*}})^2}-l_1^2}{2t_1c^{\mathrm{*}}} \\
& =\frac{\left(t_2 - t_1 \right)c^{\mathrm{*}}}{2} - \frac{1}{2c^{\mathrm{*}}}\left(\frac{l_2^2}{t_2}- \frac{l_1^2}{t_1} \right)
\\
&=\frac{\left(d_2-d_1\right)c^{\mathrm{*}}}{2c} - \frac{c}{2c^{\mathrm{*}}}\left(\frac{l_2^2}{d_2+d'}- \frac{l_1^2}{d_1+d'} \right)\,.
\end{align}
Using Pythagorean theorem $l_i^2 = d_i^2 - d'^2$ and further analytical rearrangements and simplifications lead to:
\begin{align}
\Delta x
& =\frac{\left(d_2-d_1\right)c^{\mathrm{*}}}{2c} - \frac{c}{2c^{\mathrm{*}}}\left(\frac{d_2^2 - d'^2}{d_2+d'} - \frac{d_1^2 - d'^2}{d_1+d'} \right)
\\
& = \frac{\left(d_2-d_1\right)c^{\mathrm{*}}}{2c} - \frac{c}{2c^{\mathrm{*}}}\left(d_2-d_1\right)
\\
& =  \left(\frac{d_2-d_1}{2} \right)\left(\frac{c^{\mathrm{*}}}{c}-\frac{c}{c^{\mathrm{*}}}\right)\,. \label{eq:shift_differential}
\end{align}

The above corollary formalizes the position disparity that would be observed between image frames that are beamformed with an assumed SoS $c^\mathrm{*}$ different from the 
unknown actual SoS $c_\mathrm{}$. 
Such shifts are then proportional both to a simple function of actual and beamforming SoS values, \ie  $\left(\frac{c^{\mathrm{*}}}{c_{\mathrm{}}}-\frac{c_{\mathrm{}}}{c^{\mathrm{*}}}\right)$, and also to the difference of distances of a point to Tx locations, \ie $(d_2-d_1)$.
Note that, given the utilized Tx sequences, such distance difference changes spatially across the beamforming grid.
Aggregating \eqref{eq:shift_differential} for every beamformed image point with corresponding geometric distance difference, we then arrive at the following linear system of equations:
\begin{equation} 
\label{eq:SoS_correction_linear}
\Delta{\mathbf{x}} = \mathbf{D} \gamma\ , \qquad \mathrm{where} \quad \gamma = \left(\frac{c^{\mathrm{*}}}{c_{\mathrm{}}}-\frac{c_{\mathrm{}}}{c^{\mathrm{*}}}\right)
\end{equation}
with $\Delta{\mathbf{x}}$ a vector of observed displacements, $\mathbf{D}$ a vector of differential distances, and $\gamma$ being a variable that relates assumed and actual SoS. 
When we use the actual SoS of a homogeneous medium for beamforming, there would be no shift between different beamformed frames, as expected. 
We use \eqref{eq:SoS_correction_linear} as an analytical model to estimate the beamforming SoS as follows. 

\subsection{Model fitting to estimate beamforming SoS}
Having two images beamformed with some (assumed) value, we first measure the apparent motion between these frames using a displacement tracking algorithm. 
In this work, we use time-delay estimation method based on normalized cross-correlation \cite{reza_disp}, and use this as observed $\Delta\mathbf{x}$. 
From known transmission geometry, we compute distance differences $(d_2-d_1)$ from each Tx element to every image point in the imaging grid, as $\mathbf{D}$. 
We then find the correction factor $\gamma$ by solving the inverse problem (model fitting): 
\begin{equation} 
\label{eq:SoS_correction_problem}
\hat{\gamma} = \arg\min_\gamma || \mathbf{D} \gamma - \Delta\mathbf{x} ||_2 \,.
\end{equation}
We solve \eqref{eq:SoS_correction_problem} in closed-form using $\mathbf{D}^{+}$, \ie Moore-Penrose pseudoinverse of $\mathbf{D}$, as:
\begin{equation} 
\label{eq:SoS_correction_solution_pseudoinverse}
\hat{\gamma} = \mathbf{D}^{+}\Delta\mathbf{x} \,.
\end{equation}
Using $\hat{\gamma}$ as correction factor, we then update the assumed SoS to a corrected value using:
\begin{equation} 
\label{eq:corrected_SoS_calculation}
\hat{c} = \frac{c^{\mathrm{*}} (\sqrt{\hat{\gamma}^2 +4}-\hat{\gamma})}{2}
\end{equation}

Our method can also be used iteratively. 
To do so, we first beamform the raw data using the corrected SoS value, \ie $\hat{c} \xrightarrow{}c^{\mathrm{*}}$. 
This gives more accurate beamformed frames, which leads to more accurate apparent motion observations ($\Delta\mathbf{x}$). 
Then, using these motion observations, we find a new correction factor and estimate a new corrected SoS $\hat{c}$. 
We repeat this process until corrected SoS values converges. 

\subsection{Applications on local SoS reconstruction}
\label{sec:local_sos}
Beamformed images are also a fundamental component in the reconstruction of local SoS maps, thus the global SoS assumption is implicitly essential for SoS reconstruction.
Tomographic SoS image reconstruction aims to find the heterogeneous spatial distribution of SoS within the tissue. 
For that, a differential path matrix $\mathbf{L}$ with each row indicating the integral distances from Tx and Rx elements to a spatial image location is first formed \cite{rau_2021}. 
Then, observed relative delay data $\Delta\boldsymbol{\tau}$ is a function of the spatial slowness map $\sigma$ (inverse of SoS) and its value $\sigma_0$ assumed during beamforming as $\Delta\boldsymbol{\tau} = \mathbf{L} (\sigma - \sigma_0)$.
Given observations, a local slowness map can thus be found by solving the inverse problem \cite{rau_2021}:
\begin{equation}
    \label{eq:SoS_Image_Reconstruction}
    \hat{\sigma} = \argmin_\sigma || \mathbf{L} \sigma  -({\Delta\boldsymbol{\tau}} + \mathbf{L}\sigma_0)||_1 + \lambda ||\mathbf{R}\sigma||_1,
\end{equation}
where $\mathbf{R}$ is a regularization matrix and $\lambda$ is the trade-off parameter between data and regularization terms. 
Following \cite{rau_2021}, we use total variation regularization with anisotropically weighted spatial gradients and solve the inverse problem using the L-BFGS algorithm.
For further details on local reconstruction utilized herein, refer to \cite{rau_2021}.

\section{Experiments}
We conducted experiments both using numerical simulations and tissue-mimicking phantoms.
We simulated numerical phantoms in 2D using k-Wave toolbox \cite{k_Wave}. 
We simulated the linear transducer used in the phantom experiments below, placed on a numerical domain of size $40\times55$\,mm.
Spatial and temporal simulation resolutions were set to be $75\,\mu$m and $6.25$\,ns, respectively. 

For physical testing, we used a multi-purpose multi-tissue ultrasound phantom (CIRS model 040GSE), with a manufacturer reported homogeneous background SoS value of 1540\,m/s. 
We imaged a part of the phantom with several point targets, to also evaluate improvements in B-mode resolution as a result of correcting BF SoS.
Phantom data was acquired with the UF-760AG US system (Fukuda Denshi, Tokyo, Japan), using a linear array transducer FUT-LA385-12P with $128$ channels and $300$\,mm pitch.
Sampling frequency was 40.96\,MHz. 
The data is upsampled four times before further processing.

For both numerical simulations and phantoms, we utilized diverging waves \cite{rau_2021}, \ie for each Tx only a single transducer element is activated. 
For each Tx pulse, we used 4 half cycles with a center frequency of $f_c$$=$$5$\,MHz.
Without loss of generality, for SoS correction in this paper, we used the beamformed RF frame pair from Tx of channels 55 and 65 only.
Images are beamformed on a grid of physical size $38\times50$\,mm.
We study errors due to over- and under-assumption of initial beamforming SoS values, for a range of small to large global SoS assumption errors (\{10,\,25,\,50,\,100\}\,m/s) around a known ground-truth SoS.
We study the accuracy of our corrected SoS values, as well as the benefits of such correction in B-mode imaging and local SoS reconstruction. 

\section{Numerical simulation results}
\subsection{Global SoS estimation accuracy in homogeneous medium}

\begin{figure}
\centering
\begin{tabular}{@{}cc@{}}
\includegraphics[width=0.48\linewidth]{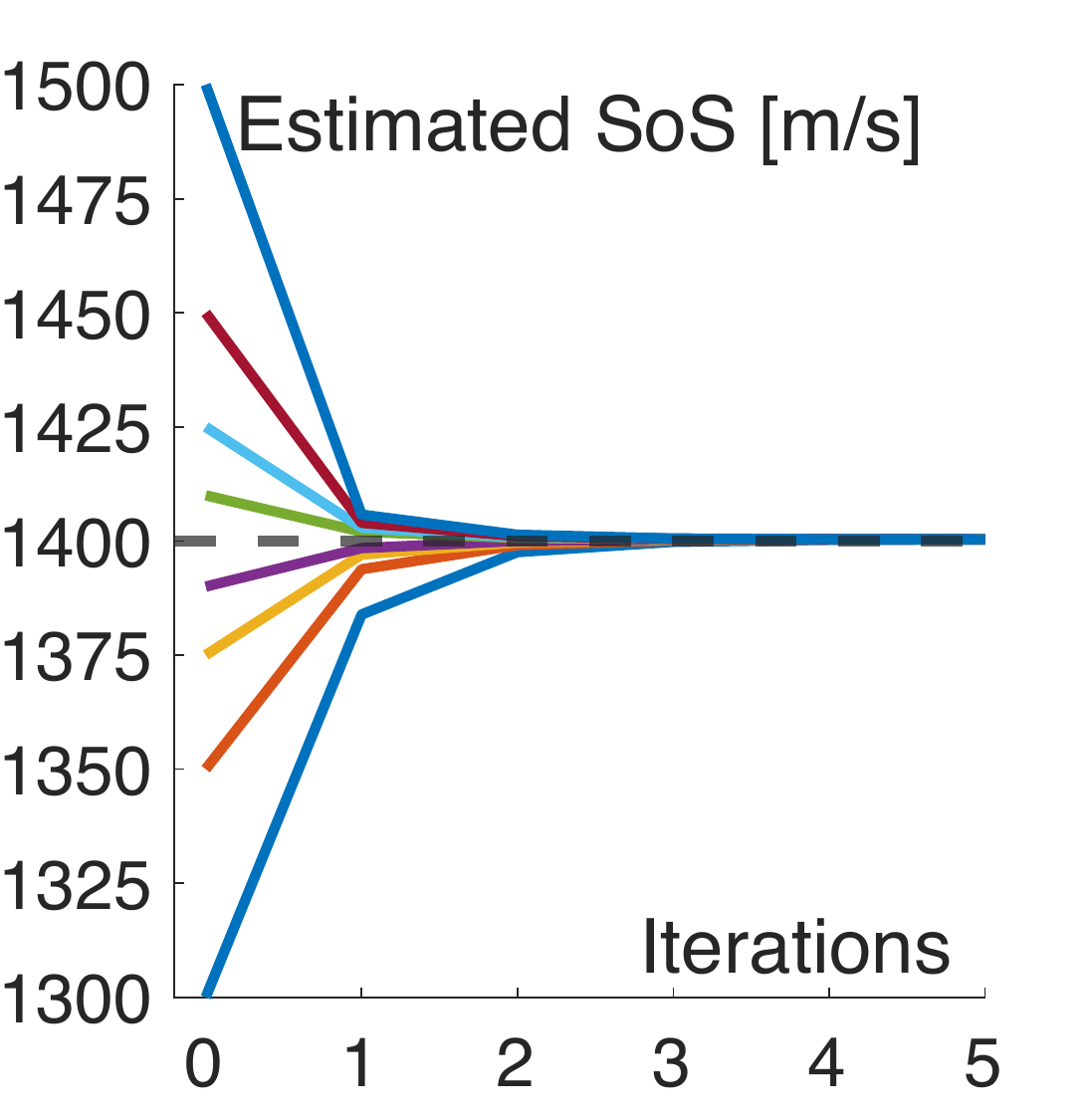}   &  \includegraphics[width=0.48\linewidth]{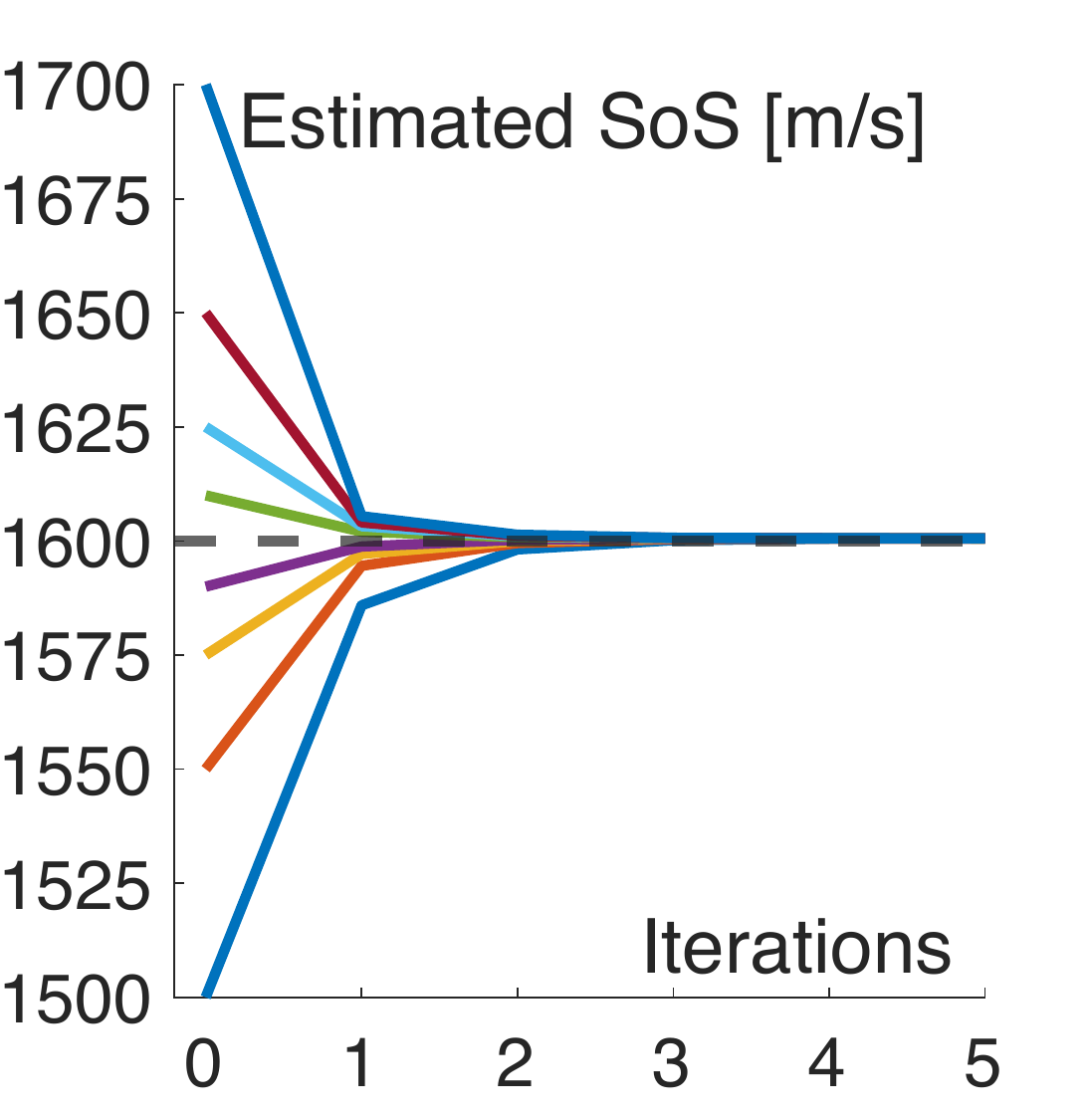} \\ 
        (a) & (b) 
    \end{tabular}
\caption{Estimation of beamforming SoS for homogeneous mediums with SoS values of (a)~1400 and (b)~1600\,m/s, starting from different initial BF SoS assumptions.}
        \label{fig:convergence_homogenous}
    \end{figure}

First, we validate our method in two homogeneous mediums with SoS values of 1400\,m/s and 1600\,m/s. 
We simulate the internal scattering by minor density variations.
\cref{fig:convergence_homogenous} shows the SoS values corrected by our method during 5 iterations.
Despite a wide range of initializations, our method finds the ground-truth SoS values within 0.6\,m/s of the set value, regardless of the initial SoS assumption. 
For convenience in presenting further results, we refer to our method of Estimating Global Speed with the acronym EGS.
Where relevant, the number of iterations $N$ that the method was run is indicated in subscript, i.e.\ EGS$_N$. 

\subsection{Improvements in B-mode images}
Here, we demonstrate how our global SoS estimation benefits B-mode imaging. 
To assess resolution, we place 3 rows of 7 point scatterers, each separated by 5\,mm and each row at imaging depths of \{25, 35, 45\}\,mm.
We use a homogeneous background SoS of 1500\,m/s in simulations.
First, we beamform images using assumed SoS values \{10,\,25,\,50,\,100\}\,m/s lower and higher than the set value. 
We then apply our SoS correction and re-beamform images with these corrected SoS values, to assess any improvements.

\begin{figure}
\centering
\includegraphics[width=\linewidth]{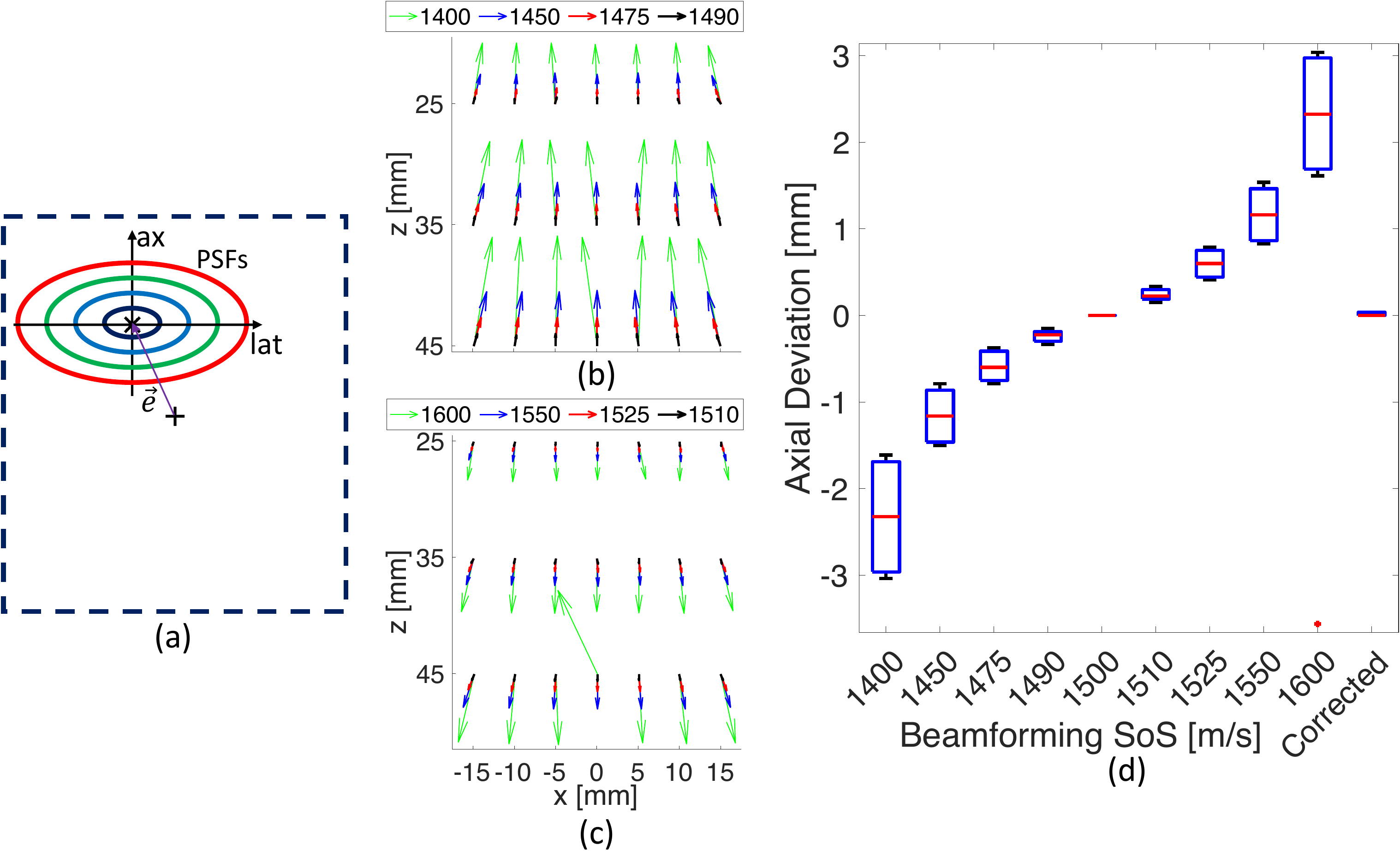}
\caption{(a)~Point scatterer evaluation procedure, showing ideal location (+), aberrated location (x), localization error ($\vec{e}$), and the lateral axis where resolution is assessed as FWHM. 
(b-c)~Localization errors in point scatterer appearances for some SoS  assumptions, with arrow lengths of 2 times physical errors for visualization. 
Corrected versions are not shown as they have almost no error.
(d)~Box plot of localization errors in individual scatterers for different SoS assumptions. }
\label{fig:bmode_arrows_localization}
\end{figure}

Initial wrong beamforming SoS assumption leads to localization artifacts, which causes point scatterers to appear at different locations in B-mode images.
To evaluate this, we first find the locations of point scatterers in beamformed images, as illustrated in \cref{fig:bmode_arrows_localization}(a). 
First, from the ideal B-mode image (obtained with SoS=1500\,m/s), we find the ground-truth location (shown with ``+'') of each point scatterer.
In each following beamformed image, we locate point scatter locations (x) by searching around (+) within a window of $10\times5$\,mm, where the spatial shift $e$ is reported as the localization error. 
Accordingly, we show sample localization errors from incorrect initial SoS assumptions in \cref{fig:bmode_arrows_localization}(b-c). 
As expected, error increases with increasing depth; and the larger the SoS assumption deviates, the larger the errors are. 
Also, largest errors from incorrect assumptions are in the axial direction.
Distribution of all axial errors with each different SoS assumption is shown in \cref{fig:bmode_arrows_localization}(d).

\begin{figure*}
\centering
\includegraphics[width=.9\linewidth]{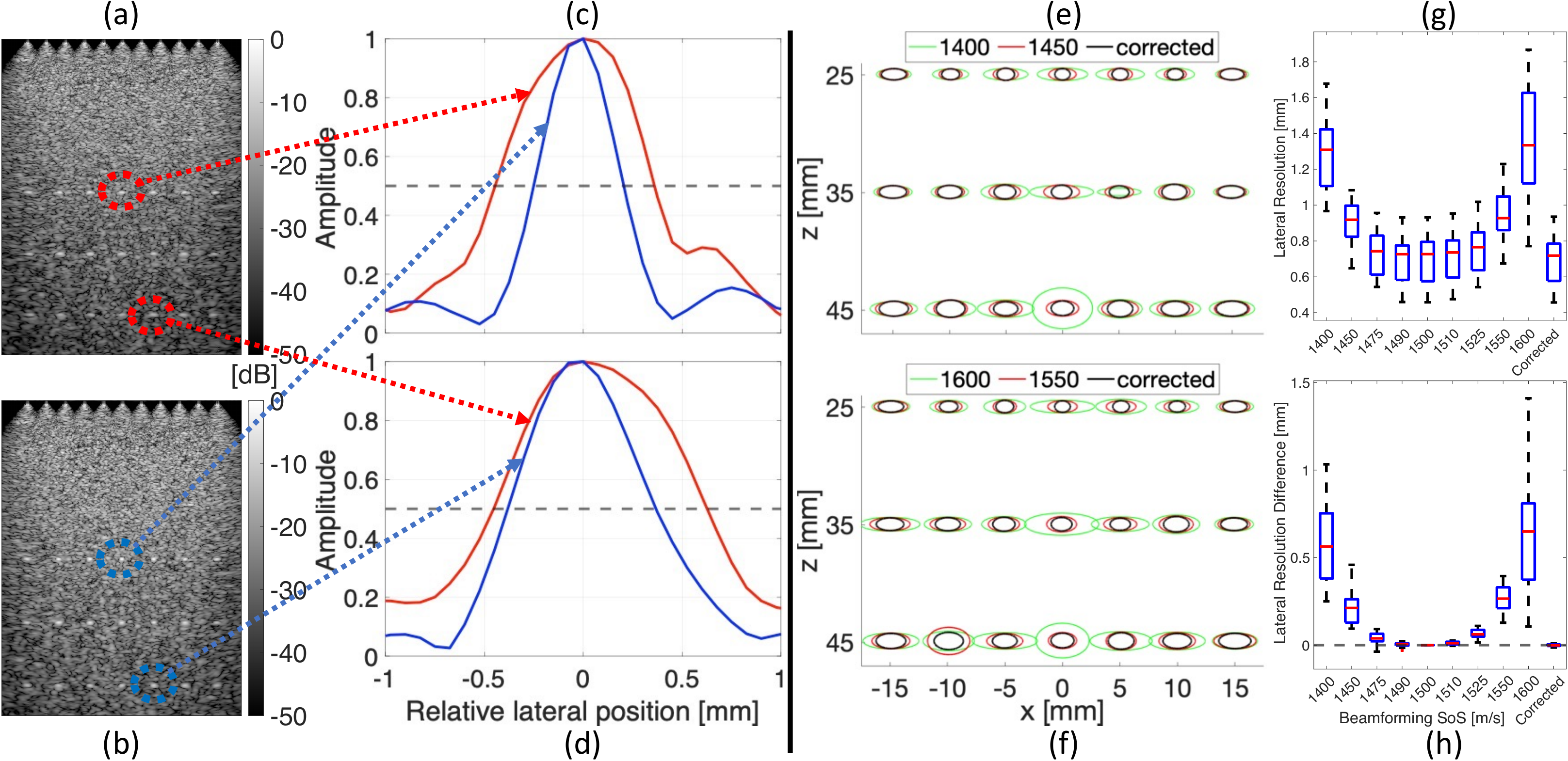}
\caption{B-mode image improvements. 
A sample experimental setting with images beamformed with (a)~assumed 1450\,m/s and (b)~its corrected SoS of 1500.6\,m/s.
Lateral envelope profiles of two sample point scatterers at (c)~middle and (d)~bottom of the image.
Illustrations of axial and lateral resolutions as ellipses for examples of (e)~under-assumption and (f)~over-assumption.
Radii of ellipses indicate the resolution in respective axes, scaled herein by 1.5 for better visibility.
(g)~Distributions of lateral resolutions for different assumed SoS values.
We show a single corrected SoS, since corrected values differ negligibly ($<$$0.04$\,m/s) with no visible effect on resolution. 
(h)~Distributions of differences of each point resolution from the ideal value of 1500\,m/s. }
\label{fig:bmode_image}
\end{figure*}

At the above-determined scatterer locations, we next quantify lateral resolution as full width at half maximum (FWHM) of the envelope signal, \ie\ from the 1D profiles along lateral axes illustrated in \cref{fig:bmode_arrows_localization}(a). 
B-mode images beamformed with assumed 1450\,m/s and its corrected value 1500.6, and corresponding sample lateral profiles are shown in \cref{fig:bmode_image}(a)-(d).
As seen, the images look sharper after our correction, and the scatterers demonstrate better resolution.
In \cref{fig:bmode_image}(e) and (f), we use ellipse plots to illustrate axial and lateral resolutions for for sample BF SoS assumptions and their corrections. 
With our corrections, the resolutions are seen to improve for all scatterers.
In \cref{fig:bmode_image}(g), we illustrate the distributions of lateral resolutions for assumed SoS values.
We also show one resolution distribution for the corrected SoS, as all corrected results show the same trend.
To normalize the natural resolution variation per depth and to assess differential errors, we present in \cref{fig:bmode_image}(h) distributions of scatterer-wise lateral resolution differences with respect to their values at ground-truth 1500\,m/s. 
As can be seen, our method improves the resolution of all points down to their ground-truth level image with known SoS. 
Lateral resolution averages tabulated in \cref{tab:improvement_in_lateral_resolution} over all 21 scatterers corroborate the above findings.
\begin{table*}
\renewcommand{\arraystretch}{1.3}
\caption{Lateral resolutions (FWHM of envelope signals) averaged over 21 point scatterers for different BF SoS assumptions and corrected values using our technique.}
\label{tab:improvement_in_lateral_resolution}
\centering
\begin{tabular}{|l|c|c|c|c|c|c|c|c|c|}
\hline
{{Initial BF SoS Assumption\,(m/s)}} & {1400}&{1450}&{1475}&{1490}&{1500}&{1510}&{1525}&{1550}&{1600}\\
\hline
{Assumed SoS\,[mm]} 
&1.33   
&0.90
&0.73
&0.69
&0.69
&0.70
&0.75
&0.95
&1.48 \\
{EGS\textsubscript{5} (proposed)\,[mm]} 
&0.69
&0.69
&0.69
&0.69
&0.69
&0.69
&0.69
&0.69
&0.69\\
\hline
{Improvement [\%]}
&48.1
&23.3
&5.5
&0
&0
&1.4
&8.0
&27.4
&53.4
\\
\hline
\end{tabular}
\end{table*}
Regardless of initial SoS assumption error, our method consistently brings lateral resolutions down to the lower-bound level of ideal BF SoS.  
For instance, sample improvements of 23.3\% and 27.4.\% are achieved for assumptions of 1450 and 1550\,m/s, respectively, corresponding to $\pm 50$\,m/s errors in BF SoS.

\subsection{Feasibility in media with heterogeneous SoS}
Next we studied the feasibility of our estimation method in media with heterogeneous SoS distributions.
We studied this using k-Wave simulations of a set of 32 numerical phantoms from \cite{melanie_2020}.
These phantoms contain mostly circular inclusions with positive and negative SoS contrast, on constant or spatially varying backgrounds; with some examples illustrated later in our result figures. 

For a heterogeneous phantom, it is not obvious what global SoS value to use as ground-truth for evaluations.
In \cite{xenia_2021,deniz2022global}, comparisons were made against the mean SoS value in the medium, however this may not be the ideal value that will best align time-delays.
For instance, the same SoS inclusion deeper or shallower on the same background would have the same image mean, but may well require different optimum global SoS values (since more beamforming paths cross the inclusion when it is closer to the transducer).
Indeed, with a global SoS setting, we wish the Rx \emph{time-delays} at each BF location to be ideal, both to individually project on the correct location and to align in-between for best resolution.
Accordingly, we set our evaluation target for heterogeneous media as the \emph{ideal} beamforming time delays at each Rx channel for each BF grid location.
We compute the ideal time-delays from the ground-truth local SoS map, using the locally-adaptive beamforming method described in \cite{rau_2018}. 
Using these as the ground-truth, BF time-delays for any other setting can be assessed, \eg as the mean absolute Time-Delay Error (TDE) across all time-delays used for the image:
\begin{equation} 
\label{eq:time_delay_error_eq}
\mathrm{TDE} = \frac{1}{N_x} \frac{1}{N_t}\sum_{m=1}^{N_x}\sum_{n=1}^{N_t}| \mathbf{t}_{m,n} -\mathbf{\hat{t}}_{m,n}|
\end{equation}
where $\mathbf{t}$ and $\mathbf{\hat{t}}$ are respectively the ideal and assessed BF time delays, $N_x$ is the number of beamforming grid points, and $N_t$ is the number of Rx channels in aperture.  

In~\cref{tab:time_delay_imp}, we present average TDE over 32 phantoms, for different
initial BF SoS assumptions and for their corrected values after the 1st and the 5th method iterations.
\begin{table*}
\caption{Mean absolute time-delay errors across the entire images, for different BF SoS assumptions and their respective corrected SoS values. }
\label{tab:time_delay_imp} 
\centering
\begin{tabular}{|l|c|c|c|c|c|c|c|c|c|}
\hline
{{Initial BF SoS Assumption\,(m/s)}} & {1400}&{1450}&{1475}&{1490}&{1500}&{1510}&{1525}&{1550}&{1600}\\
\hline
{Assumed SoS\,[{$\mathbf{\mu}$}s]} 
&1.46
&0.72
&0.37
&0.17
&0.04
&0.11
&0.30
&0.61
&1.22
\\
{EGS\textsubscript{1}(\cite{xenia_2021})\,[{$\mathbf{\mu}$}s]}
&1.01
&0.55
&0.23
&0.12
&0.10
&0.10
&0.24
&0.49
&0.97
\\
{EGS\textsubscript{5}(\cite{xenia_2021})\,[{$\mathbf{\mu}$}s]}
&0.36
&0.16
&0.09
&0.08
&0.08
&0.08
&0.06
&0.11
&0.34
\\
{EGS\textsubscript{1}\,[{$\mathbf{\mu}$}s]}
&0.26
&0.10
&0.09
&0.07
&0.06
&0.06
&0.06
&0.07
&0.10
\\
\textbf{EGS\textsubscript{5} (proposed)\,[{$\boldsymbol{\mu}$}s]}
&0.07
&0.07
&0.07
&0.07
&0.07
&0.07
&0.07
&0.07
&0.07
\\
\hline
{Improvement [$\times$\,folds]}
&20.9
&10.3
&5.3
&2.4   
&0.6
&1.6
&4.3
&8.7
&17.4
\\
\hline
\end{tabular}
\end{table*}
For comparison, we also computed corrections by replacing the geometrical model \eqref{eq:shift_differential} in our proposed method, with the model $\left(\frac{2}{c}-\frac{2}{c^\mathrm{*}}\right)$ of \cite{xenia_2021}.
Note that the latter model is a simplified approximation proposed in \cite{xenia_2021} in order to guide model fitting, for which any residual errors were corrected using a calibration process.
Presently in this work, since our improved and accurate geometric model directly relates observations to global SoS, we do not implement nor need any calibration process.

\begin{figure*}
      \centering
      \subfigure[TDE\,{[$\mu$s]}] {\includegraphics[height =0.26 \linewidth]{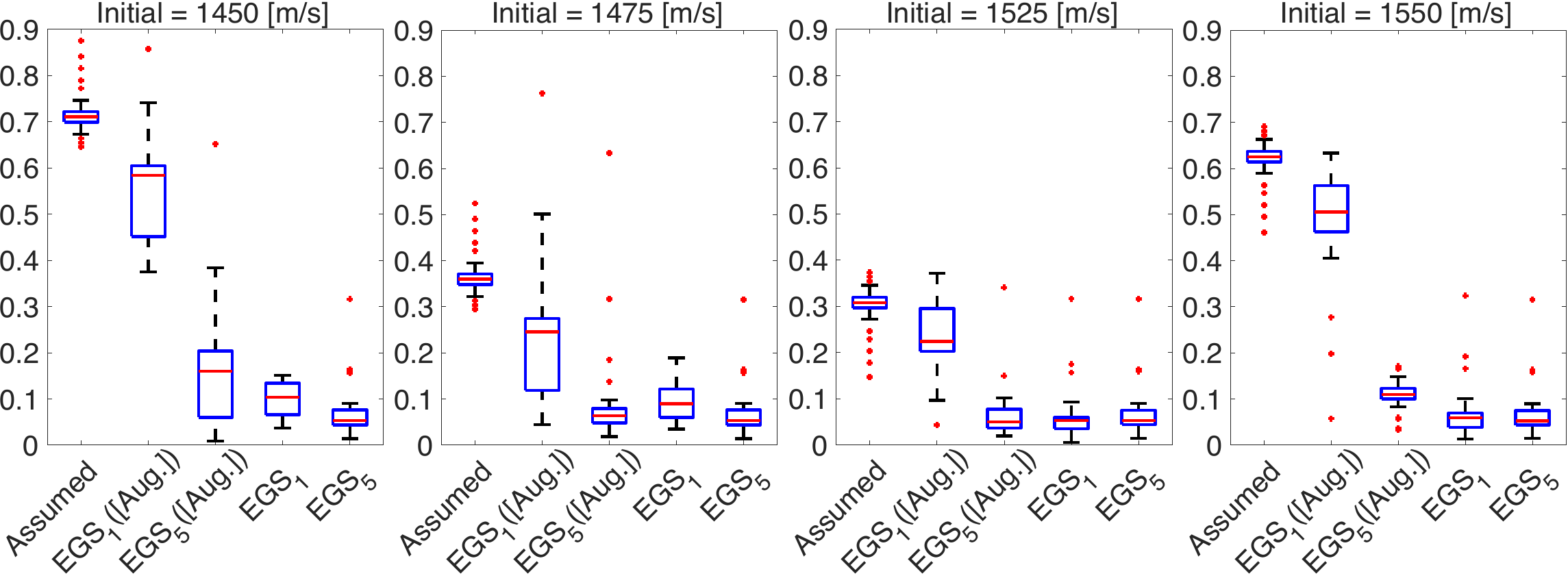}}  \hspace{0.2em}
         \subfigure[RMSE\,{[m/s]}] {\includegraphics[height =0.26 \linewidth]{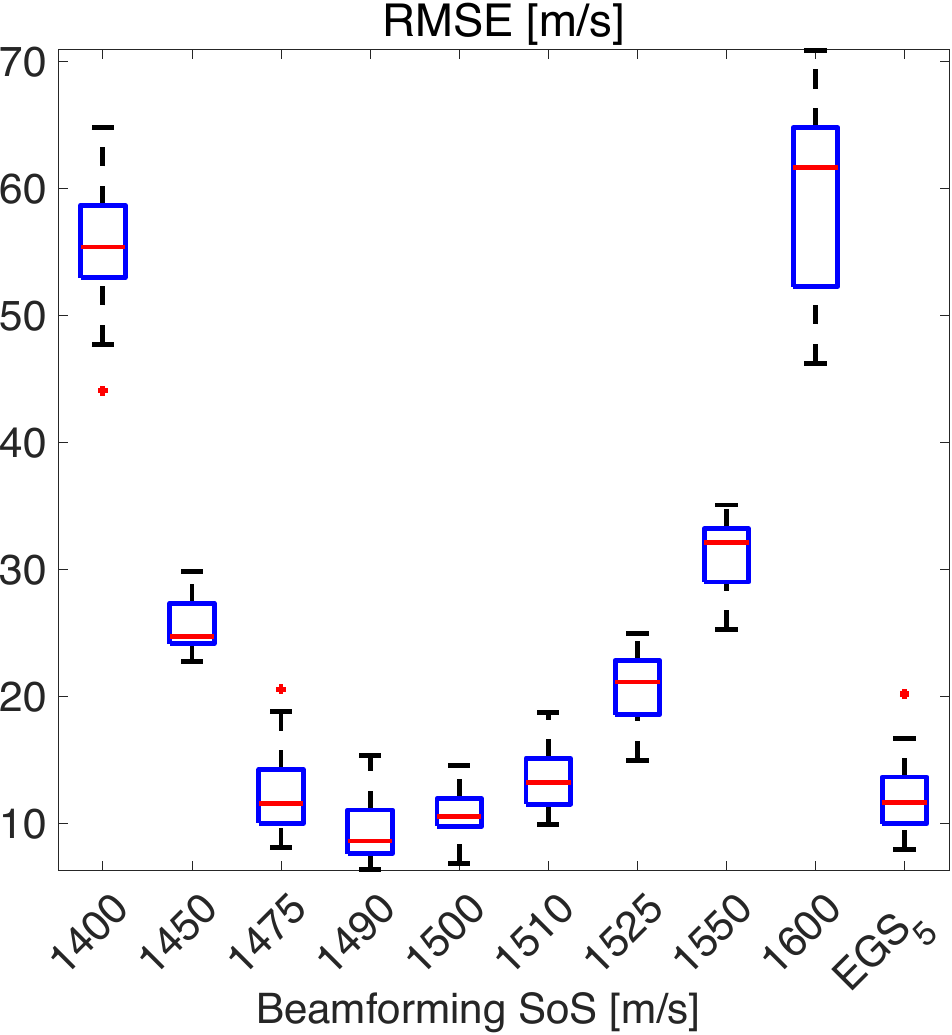}}  \hspace{0.3em} 
        \caption{(a)~Time-delay error (TDE) distributions per phantom using different methods, for sample SoS assumptions. (b)~Root mean square error (RMSE) distributions for all SoS assumptions.
        A single corrected SoS is shown, since corrected values of a phantom from different initializations differ negligibly ($<$$0.4$\,m/s).
        EGS([Aug.]) refers to using our method with (\eqref{eq:shift_differential}) replaced with the estimation model proposed in~\cite{xenia_2021}.}
        \label{fig:box_plot_delay_us_and_SoS_boxp}
    \end{figure*}

As seen in \cref{tab:time_delay_imp}, our method significantly reduces time-delay errors compared to incorrect assumptions of initial SoS values. 
Improvements over the geometric model of \cite{xenia_2021} are also substantial, even when using an iterative scheme, especially for larger initial BF SoS deviations.
In~\cref{fig:box_plot_delay_us_and_SoS_boxp}(a), distributions per phantom for four experimental settings ($\pm$\{25,50\}\,m/s initial offset) help illustrate the TDE accuracy improvements attained by our proposed method. 
Our method effectively minimizes TDE across the whole image.
Note that for heterogeneous phantoms, TDE cannot vanish since local SoS variations cannot be faithfully modeled by a single global SoS \cite{rau_2018}.
Strikingly, our method finds the same global SoS values from any initialization utilized within a wide range, as can be seen in~\cref{tab:time_delay_imp} and~\cref{fig:box_plot_delay_us_and_SoS_boxp}(a).
    
\subsection{Impact on local SoS reconstruction}
\begin{figure*}
\centering
\includegraphics[width=\linewidth]{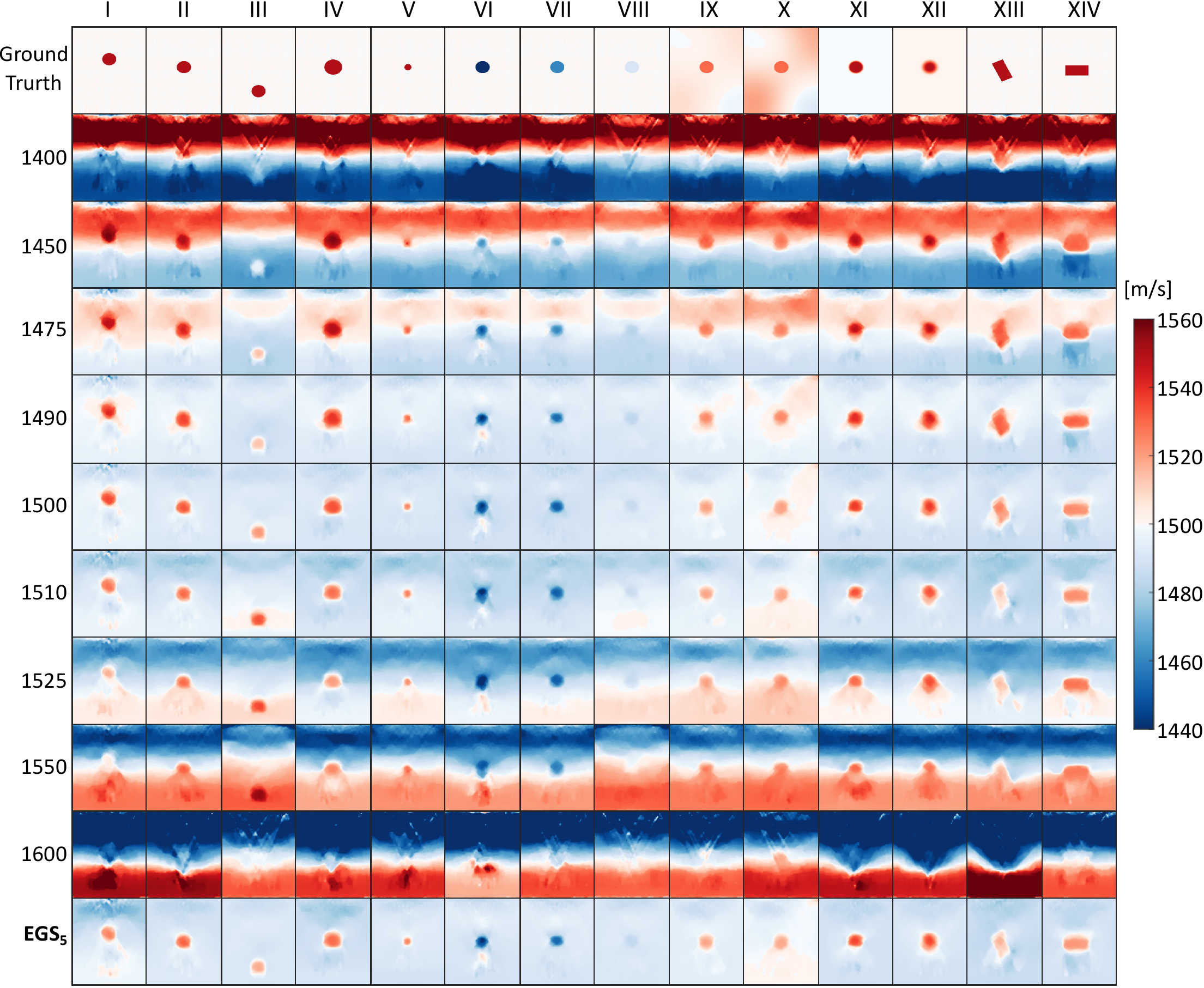}
\caption{Example SoS reconstructions for sample numerical phantoms (top row) with inclusions of varying sizes, shapes, and SoS contrasts. Reconstructions are obtained using different BF SoS assumptions and our corrected values (reported on the left).}
\label{fig:simulation_sos_recons}
\end{figure*}

Local SoS reconstruction uses displacements between beamformed RF frames, thus its accuracy also relies on good beamforming and hence a correct global SoS assumption.
To study this, we compared local SoS reconstructions of the 32 phantoms, using under-/over-assumed beamforming SoS values as well as values corrected by different methods. 
For local SoS reconstruction, we used 6 pairs of diverging wave transmits as in \cite{melanie_2020} and adaptive receive aperture \cite{rau_2021}.
We evaluated each reconstruction via root mean square error (RMSE) with respect to the known ground truth local SoS map, as:
\begin{equation} 
\label{eq:RMSE_form}
\mathrm{RMSE}(\mathbf{x},\mathbf{\hat{x}}) = \sqrt{\frac{1}{N_x}\sum_{n=1}^{N_x}\left(\mathbf{x}_n -\mathbf{\hat{x}}_n\right)^2},
\end{equation}
where $\mathbf{x}$ and $\mathbf{\hat{x}}$ are respectively the ground truth and reconstructed SoS distributions, and $N_x$ is the number of pixels in the reconstructions. 
Distributions of RMSEs for 32 numerical phantoms shown in \cref{fig:box_plot_delay_us_and_SoS_boxp}(b) and their average values reported in~\cref{tab:RMSE_values}. 
\begin{table*}
\renewcommand{\arraystretch}{1.3}
\caption{Root mean square errors of local SoS reconstructions, averaged over 32 numerical phantoms using different BF SoS assumptions and their corrected SoS values.}
\label{tab:RMSE_values}
\centering
\begin{tabular}{|l|c|c|c|c|c|c|c|c|c|}
\hline
{{Initial BF SoS Assumption\,[m/s]}} & {1400}&{1450}&{1475}&{1490}&{1500}&{1510}&{1525}&{1550}&{1600}\\
\hline
{Assumed SoS\,[m/s]} 
&55.55
&25.73
&12.27
&9.36
&10.81
&13.44
&20.63
&31.19
&58.86 \\
{EGS\textsubscript{5} (proposed)\,[m/s]} 
&12.09
&12.10
&12.11
&12.10
&12.11
&12.10
&12.11
&12.12
&12.12\\
\hline
{Improvement [$\times$\,folds]}
&4.6
&2.1
&1.0
&0.8   
&0.9
&1.1
&1.7
&2.6
&4.9
\\
\hline
\end{tabular}
\end{table*}
\cref{tab:RMSE_values} show that our proposed method substantially improves the accuracy of tomographic SoS reconstructions, especially for larger errors in SoS assumption.
For instance, for beamforming SoS values of 1450 and 1550\,m/s, RMSE is improved by 2.1 and 2.6 folds, respectively.
Similarly, for beamforming SoS values of 1400 and 1600\,m/s, RMSE is improved by 4.6 and 4.0 folds.
Importantly, our method makes the reconstruction outcomes independent of and robust to the initial setting of the unknown global SoS value for beamforming.
We show reconstructions of 14 sample simulation inclusions in \cref{fig:simulation_sos_recons}, with different assumed BF SoS values as well as with our corrected values. 
As mentioned above and seen in the earlier result tables, thanks to the robustness of our technique EGS, the corrected global SoS values from different initial assumptions are very similar, which in turn makes the corrected tomographic reconstructions relatively similar.
Accordingly, we herein present only one corrected reconstruction, from the extreme initial BF SoS of 1600\,m/s.
As seen, our global SoS estimation method EGS allows for overall improved tomographic SoS reconstructions.
More importantly, our method enables optimal reconstructions \emph{independent} of unknown medium SoS.
For instance, the reconstruction results with assumed values 1450 and 1550\,m/s, which are around 3\% range of expectable variations within tissue, are severely subpar to the reconstructions using our corrected values.

\section{Phantom results}

\subsection{Global SoS estimation}

\begin{figure}
\centering
\includegraphics[width=0.5\linewidth]{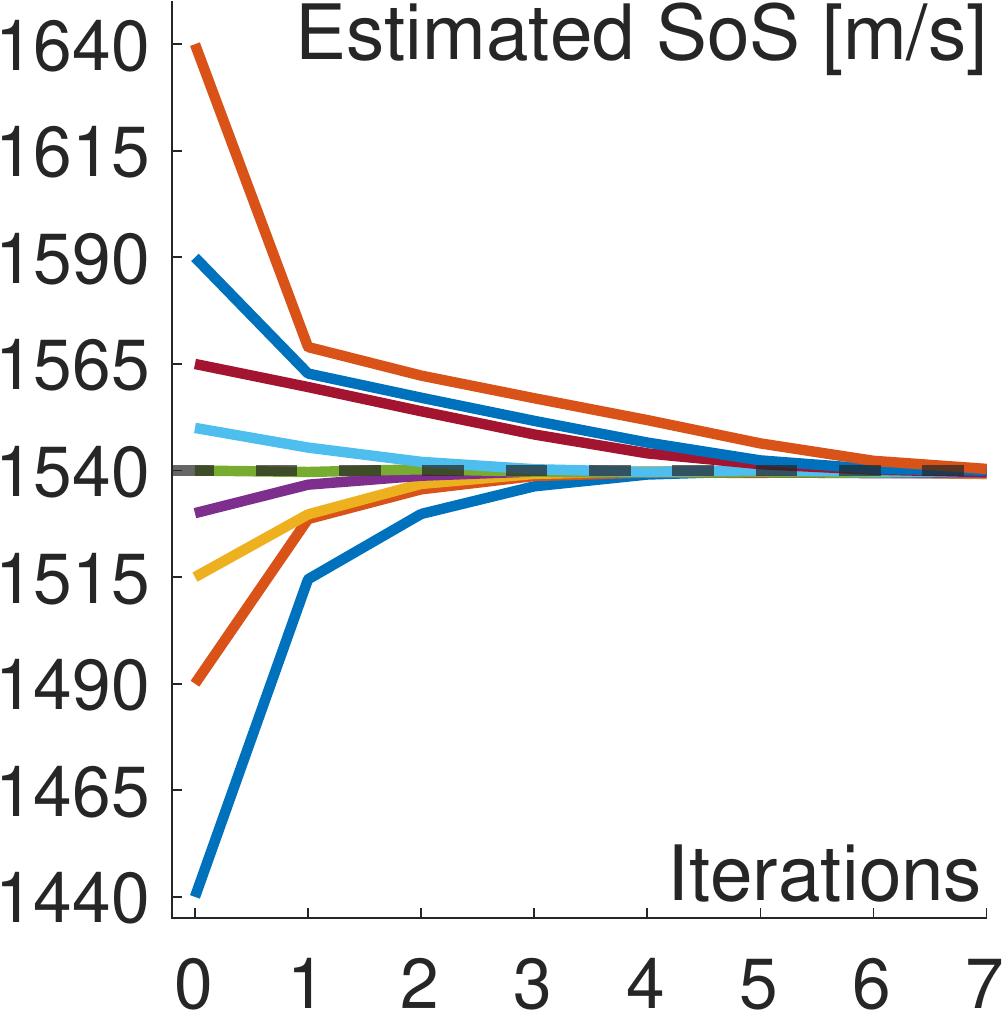}
\caption{Iterative estimation of BF SoS for a tissue-mimicking phantom with known SoS value of 1540\,m/s, starting from different initial BF SoS assumptions. }
\label{fig:homogenous_convergence}
\end{figure}

On the data we collected from the CIRS phantom, we apply our method starting from 8 different initial SoS assumptions.
Iterative estimation of corrected global SoS values is shown in \cref{fig:homogenous_convergence}, where the manufacturer reported value of 1540\,m/s is converged within 1\,m/s in 7 iterations.

\subsection{Improvements in B-mode images}
Next we studied the effect of correcting global SoS on B-mode image quality, similarly to the numerical experiments.
We computed the lateral resolutions of 7 point targets within the imaged field-of-view, both with the assumed and our corrected BF SoS values.
Average lateral resolutions tabulated in \cref{tab:improvement_in_lateral_resolution_exp_data} indicate that B-mode image quality is improved in all cases, with resolution improvements exceeding 50\% for extreme cases of erroneous assumptions. 
\begin{table*}
\caption{Lateral resolutions (FWHM of envelope signals) averaged over 7 point scatterers for different BF SoS assumptions and their corrected values.}
\label{tab:improvement_in_lateral_resolution_exp_data}
\centering
\begin{tabular}{|l|c|c|c|c|c|c|c|c|c|}
\hline
{{Initial BF SoS Assumption [m/s]}} & {1440}&{1490}&{1515}&{1530}&{1540}&{1550}&{1565}&{1590}&{1640}\\
\hline
{Assumed SoS\,[mm]} 
&1.22 
&0.81
&0.62
&0.57
&0.56
&0.56
&0.60
&0.73
&1.16 \\
{EGS\textsubscript{7}\, (proposed)\,[mm]} 
&0.56
&0.56
&0.56
&0.56
&0.56
&0.56
&0.56
&0.56
&0.56\\
\hline
{Improvement\,[\%]}
&54.1
&30.9
&9.7
&1.8
&0
&0
&6.7
&23.3
&51.7
\\
\hline
\end{tabular}
\end{table*}
\begin{figure*}
\centering
\includegraphics[width=.95\linewidth]{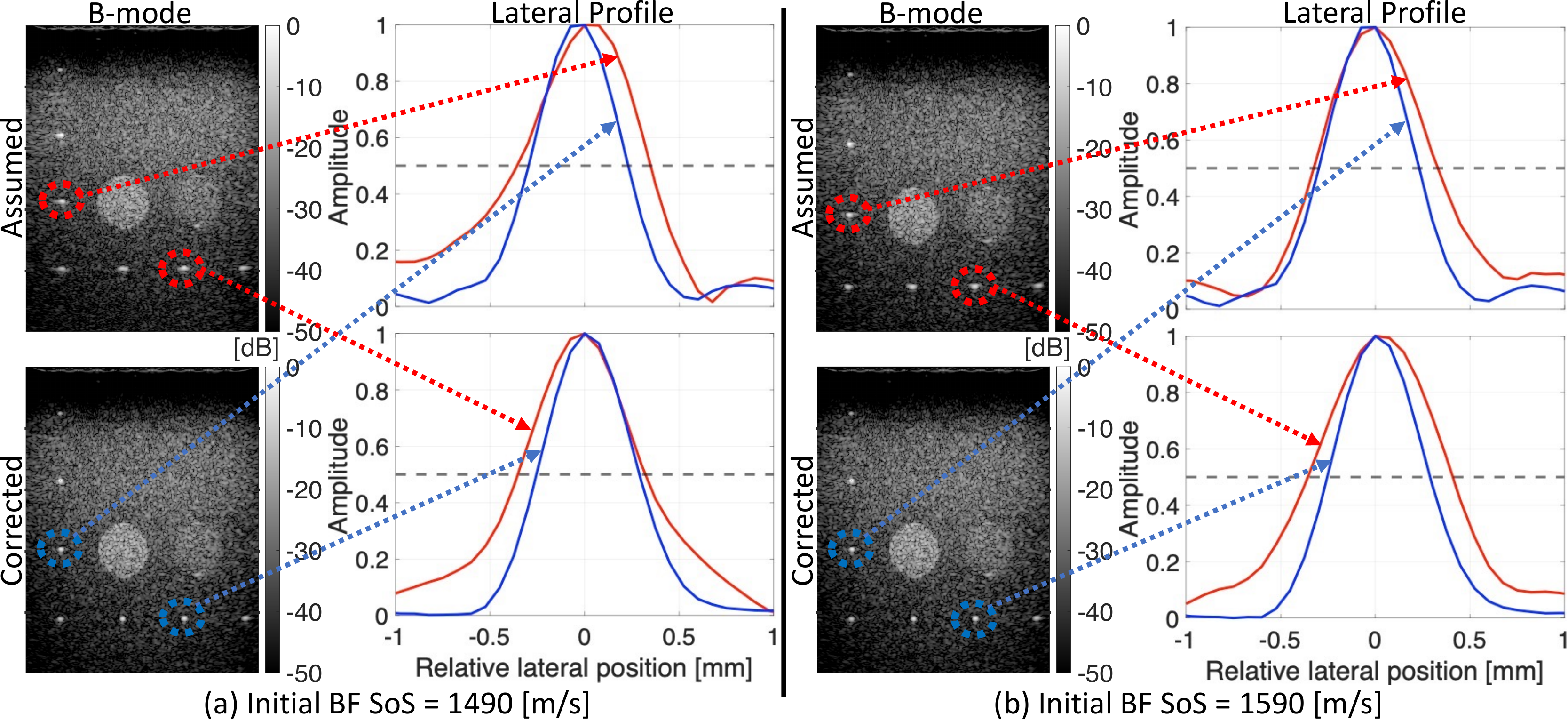}
\caption{B-mode improvements in phantoms, with 
results shown for two sample experimental settings with assumed BF SoS of (a)~1490\,m/s and (b)~1590\,m/s.
Corresponding corrected values are 1540.1 and 1540.2\,m/s, respectively.
B-mode images demonstrate the substantial improvements in image sharpness and point localization with our corrections.
Lateral envelope profiles for two sample scatterers in all shown settings corroborate the observations.
}
\label{fig:bmode_image_experimental}
\end{figure*}
In~\cref{fig:bmode_image_experimental} we show the B-mode resolution improvements from two sample assumed values of 1490 and 1590\,m/s, each image showing lateral profiles of two sample scatterer. 
As seen, the blurrier appearances of points in B-mode become sharper after our correction, as well visible from the reduced FWHM of their lateral profiles. 
As can be seen with respect to the axis markers, the mislocalization of scatterers are also corrected using out technique.

\section{Discussion and Conclusions}
In this work, we have presented an analytical method to estimate global BF SoS using geometric disparities between different Tx events. 
This method utilizes the known transducer geometry, and does not need any training or a-priori calibration.
We have studied our method on numerical simulations and experimental phantoms, and illustrated its utility in improving B-mode image resolution and correcting localization errors.
With numerical simulations of heterogeneous SoS, we showed our method to improve the accuracy of both beamforming time delays and local SoS reconstructions.
Our proposed method is robust to different SoS initializations and generalizable to various test settings.

Although we utilized diverging waves herein, one could also use other Tx sequences, \eg plane waves from differently (but nonsymmetric) angled directions.
However, with planes waves, the geometric disparities would exist only in the axial direction, which could make it difficult to estimate the global SoS value, due to the likely confounding effects of the inherent axial SoS variations across typically layered tissue structures.
For diverging waves, geometric disparities between different Tx events change laterally to a large extent.
As shown in our results with heterogeneous numerical phantoms, we can estimate beamforming SoS robustly, regardless of initial error.  
We hope to test the method in in-vivo settings in future studies.

Our SoS correction method can be readily integrated into real-time imaging settings.
Note that $\mathbf{D}^{+}$ for a fixed imaging depth is a predefined constant matrix, which can be precomputed.
Then, estimation in closed-form only requires a vector inner-product, which can be computed very fast.
Iterative application of our method requires beamforming at each step, which may be a computational bottleneck.
Nevertheless, one can run iterative steps described herein for consecutive frames, while beamforming each with the corrected value from the previous.
Since tissue scenes in ultrasound exams do not change rapidly, a correct value can be found within a few frames at a new location, after which only differential SoS updates are necessitated.

Our method can also be beneficial in shear-wave elastography, as beamforming SoS was shown in~\cite{Chintada_phase-aberration_21} to affect shear-wave speed estimation.
Furthermore, since SoS can be a pathological indication, diagnostic applications can also be envisioned.
Our global method could potentially also be extended to a local SoS estimator, e.g.\ through differentiation of applying it at multiple depths.
Future work will focus on testing some of these directions.

\bibliographystyle{IEEEtran}
\bibliography{sample}
\end{document}